\documentclass[graybox]{svmult}

\usepackage{mathptmx}       
\usepackage{helvet}         
\usepackage{courier}        
\usepackage{type1cm}        
\usepackage{makeidx}         
\usepackage{graphicx}        
\usepackage{multicol}        
\usepackage[bottom]{footmisc}

\makeindex             

\begin{document}

\title*{Metallicities of galaxies in the nearby Lynx-Cancer void}
\author{Alexei Kniazev, Simon Pustilnik, Arina Tepliakova and Alexander Burenkov}
\institute{Alexei Kniazev \at SAAO, PO Box 9, 7935 Observatory, Cape Town, South Africa, \email{akniazev@saao.ac.za}
\and Simon Pustilnik, Arina Tepliakova and Alexander Burenkov \at SAO, Nizhnij Arkhyz, Karachai-Circassia, 369167, Russia, \email{sap,arina,ban@sao.ru}}
\maketitle

\vskip-1.2truein

\abstract{
Does the void environment have a sizable effect on the evolution of dwarf
galaxies? If yes, the best probes should be the most fragile least massive
dwarfs. We compiled a sample of about one hundred dwarfs with M$_{\rm B}$
in the range --12 to --18 mag, falling within the nearby Lynx-Cancer
void. The goal is to study their evolutionary parameters -- gas metallicity
and gas mass-fraction, and to address the epoch of the first substantial
episode of Star Formation. Here we present and discuss the results of O/H
measurements in 38 void galaxies, among which several the most metal-poor
galaxies are found with the oxygen abundances of 12+log(O/H)=7.12-7.3 dex.}

\section{Objectives}
\label{sec:1}
In the framework of Cold Dark Matter models, dwarf galaxies in voids
could form later and evolve more slowly than their counterparts in a more
typical environment. However, quantitative predictions are uncertain.
Data on evolution of void population are rather scarce and indirect.
We recently described a nearby void in Lynx-Cancer (Pustilnik \& Tepliakova,
2010, MNRAS, submitted) with D$_{\rm cent}$ $\sim$18~Mpc and the size
$\sim$16~Mpc. About 100 dwarfs fall in this void. The goal of the ongoing
project is to obtain the evolutionary parameters of the void sample galaxies.
Here we present the intermediate results of O/H determination, based
mainly on the SAO 6-m telescope observations.

\begin{figure}[b]
\sidecaption
\includegraphics[width=7.5cm]{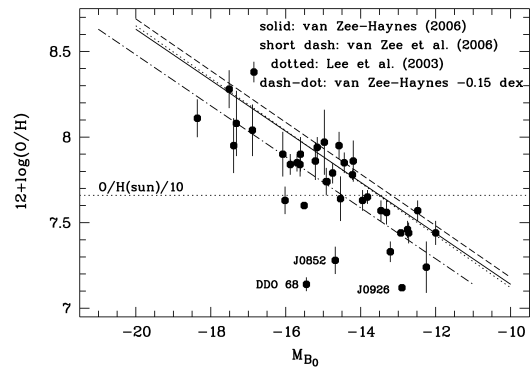}
\caption{Metallicity-Luminosity relation for 38 Lynx-Cancer void galaxies.
Solid, short-dash and dotted lines show respectively the known L-Z relations
for isolated late-type galaxies (\cite{vZH06}), nearby dIs (\cite{vZ06}) and
nearby dI and I galaxies from \cite{l03}. The dash-dot line is 0.15 dex lower
than the solid one (for which the fit rms=0.15 dex). Many \mbox{L-C} void
galaxies (especially those with M$_{\rm B}> -16.0$) fall below the dash-dot
line.}
\label{fig:Kniazev1}
\end{figure}

\section{Results}
\label{sec:2}
Currently we have oxygen abundances for 38 Lynx-Cancer void galaxies,
which are shown as the diagram O/H versus M$_{\rm B}$ in
Fig.~\ref{fig:Kniazev1}. The majority of galaxies have O/H derived via classic
T$_{\rm e}$-method with programs described in~\cite{K08}.  About 1/3 of
dwarfs with faint or undetected [OIII] $\lambda$4363 lines have O/H derived
via the semi-empirical method of~\cite{IT07}. Three lines, close to each
other, show fits to similar empirical relations for local I/dI galaxies and
isolated late-type galaxies from \cite{l03,vZH06}
and \cite{vZ06}.
The dash-dot line is just shifted down by 0.15 dex relative to that by
\cite{vZH06}. About 1/3 of void galaxies have O/H below this
line. The effect looks more prominent for M$_{\rm B} > -16$.
The latter could be the first indication of the slower chemical evolution of
the least massive galaxies in voids.

\section{Unusual dwarfs in Lynx-Cancer void}
\label{sec:3}
In the course of this void galaxy sample study, an unusual concentration
of the most metal-poor dwarfs is found. Namely, at least 5 galaxies with
12+log(O/H)$\le$7.3 fall in this void: J0926+3343 (7.12) \cite{J0926},
DDO 68 (7.14) \cite{DDO68,IT07}, J0737+4724 (7.24), J0852+1350 (7.28),
J0812+4836 (7.28) \cite{IT07}.
In contrast to the majority of the void galaxies, the former three have SDSS
colours in their outer parts, indicating no traces of stars with ages
more than 1--3 Gyr. Two more Lynx-Cancer void LSBDs, SAO 0822+3545 and SDSS
J0723+3622, show no traces of their older stellar population.
The void sample shows the sizable overabundance of the most metal-poor
objects compared to the Local Volume sample.
This fact also suggests slower chemical evolution of void dwarfs.

\end{document}